\documentclass[preprint]{elsarticle}  

\usepackage{hyperref}

\journal{...}

\usepackage{numcompress}\biboptions{authoryear}
\usepackage{amsfonts}
\usepackage{amsmath} 
\usepackage{bm}
\usepackage{xcolor, soul}
\sethlcolor{green}

\usepackage{hyperref}
\usepackage{breakurl}

\usepackage{graphicx}

\newcommand{\xM}{{\bm x}}
\newcommand{\xT}{{\bm y}}
\newcommand{\pxM}{p\left(\xM\right)}
\newcommand{\pxT}{p\left(\xT\right)}
\newcommand{\pxMT}{p\left(\xM\right.\left|\,\xT\right)}
\renewcommand{\hat}{\widehat}

\begin{document}

\makeatletter
\def\ps@pprintTitle{%
  \let\@oddhead\@empty
  \let\@evenhead\@empty
  \def\@oddfoot{\reset@font\hfil\thepage\hfil}
  \let\@evenfoot\@oddfoot
}
\makeatother

\begin{frontmatter}

\title{Estimation of circular statistics in the presence of measurement bias}

\author[unmc]{Abdallah Alsammani\fnref{nowat}}
\author[umBME,umN]{William C. Stacey}
\author[unmc]{Stephen V. Gliske\corref{cor1}}
\cortext[cor1]{Corresponding author}
\ead{steve.gliske@unmc.edu}
\fntext[nowat]{Now at the School of Science and Mathematics, Jacksonville University, Jacksonville, Florida}

\address[unmc]{Department of Neurosurgery, University of Nebraska Medical Center, Omaha, Nebraska}
\address[umBME]{Department of Biomedical Engineering, University of Michigan, Ann Arbor, Michigan}
\address[umN]{Department of Neurology, University of Michigan, Ann Arbor, Michigan}

\begin{abstract}
  \paragraph{Background and objective}
  Circular statistics and Rayleigh tests are important tools for analyzing the occurrence of cyclic events. However, current methods fail in the presence of measurement bias, such as incomplete or otherwise non-uniform sampling. Consider, for example, studying 24-cyclicity but having data not recorded uniformly over the full 24-hour cycle. The objective of this paper is to present a method to estimate circular statistics and their statistical significance even in this circumstance.

  \paragraph{Methods}
  We present our objective as a special case of a more general problem: estimating probability distributions in the context of imperfect measurements, a highly studied problem in high energy physics. Our solution combines 1) existing approaches that estimate the measurement process via numeric simulation and 2) innovative use of linear parametrizations of the underlying distributions. We compute the estimation error for several toy examples as well as a real-world example: analyzing the 24-hour cyclicity of an electrographic biomarker of epileptic tissue controlled for state of vigilance.
    
  \paragraph{Results} 
  Our method shows low estimation error. In a real-world example, we observed the corrected moments had a root mean square residual less than 0.007. We additionally found that, even with unfolding, Rayleigh test statistics still often underestimate the p-values (and thus overestimate statistical significance) in the presence of non-uniform sampling. Numerical estimation of statistical significance, as described herein, is thus preferable.
     
  \paragraph{Conclusions}
  The presented methods provide a robust solution to addressing incomplete or otherwise non-uniform sampling. The general method presented is also applicable to a wider set of analyses involving estimation of the true probability distribution adjusted for imperfect measurement processes.

\paragraph{Highlights}
\begin{itemize}
\item Circular statistics and Fourier expansions are directly related.
\item Methods to address non-uniform sampling are widely used in high energy physics.
\item Basis function expansions address the limitations of prior methods.
\item Our solution has very low estimation errors.
\item Numeric estimation, not Rayleigh tests, should be used to assess statistical significance.
\end{itemize}

\end{abstract}


\end{frontmatter}


\section{Introduction}
The analysis of cyclic events is part of the branch of statics called
circular, directional, or spherical statistics. One main application in
biomedical research is the analysis of events whose occurrence rate
varies with circadian rhythm.  For example, circular static moments
were used to analyze the daily and multiday cycling of seizures in
individuals with epilepsy \citep{baud2018,karoly2018}.  However, the
standard methods of computing circular statistics fail to account for
non-uniform and incomplete sampling.  By non-uniform or incomplete
sampling, we refer to limitations of the measurement process in which
the sampling from various regions of the domain is not
consistent. Incomplete sampling is one type of non-uniform sampling.
For example, if one only records events during a subset of the 24-hour
period but wants to understand the 24-hour cyclicity of these events,
they will have incomplete and non-uniform sampling.  If instead, one
records over the full 24-hour period but records from the morning
hours twice as often as from the afternoon, they will have complete,
but still non-uniform, sampling. Non-uniform sampling of any type,
including incomplete sampling, can bias the estimation of the
cyclicity of these events.

\begin{figure}
  \begin{center}
  \includegraphics[width=\textwidth]{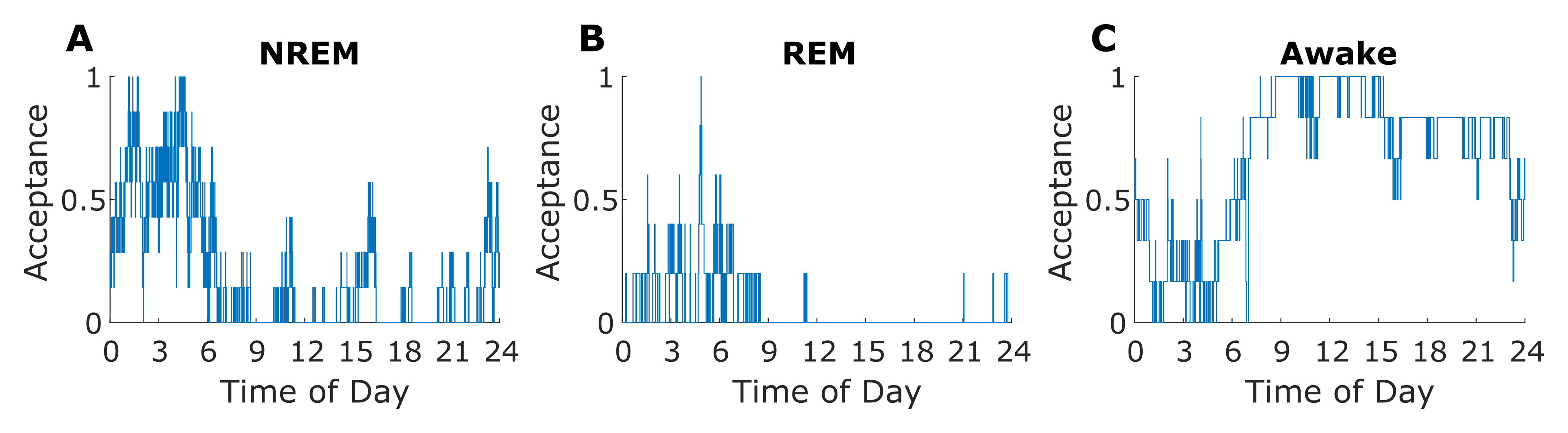}
  \caption{{\bf Example distributions of state of vigilance versus time of day.} Data shown for an example patient in states of vigilance NREM (A), REM (B), and awake (C).  Notice the highly irregular
    sampling, with nearly all REM data coming from hours 0 to 9 and
    the majority of awake data coming from hours 9 to 24.  The y-scale is set so that the maximum value is 1.0.  }
  \label{fig:example}
  \end{center}
\end{figure}

To give a more realistic example, our research group has been actively involved in the analysis of biomarkers of epilepsy that occur in intracranial EEG recordings.  We have multiday recordings of hospitalized subjects with epilepsy, and these subjects have highly irregular sleep patterns.  We would like to assess if the rate of certain biomarkers has circadian oscillations that are independent of the sleep stage.  One approach is to stratify the data by sleep stage and then to compute the first circular statistic moment for each sleep stage.  This involves pooling the data across the multiple day recordings.  Plotting the frequency of sleep stages versus time shows that our data has highly non-uniform sampling; see Fig. \ref{fig:example}.  Thus, methods are needed to address the impact of this high level of non-uniform and incomplete sampling.

Correcting for non-uniform sampling
is common in spin physics, a sub-field of high energy nuclear
physics. In that case, the measured angular distributions are related
to the underlying distribution of angular momentum of quarks and
gluons and the non-uniform sampling is due to the shape and other
physical characteristics of the particle physics detector.  In spin
physics, the angular distributions are also depend on multiple, additional
variables, denoted kinematic variables.  In biomedical research, these
additional variables would be called confounding factors, and in the
context of circadian rhythm, the effect of these confounding factors
is called masking \citep{klerman2017}.

Techniques to correct for non-uniform and incomplete sampling are extremely common in high energy physics. Most
often, these techniques are based on creating histograms of
counts \citep{hoecker1996}. For example, see
\citet{airapetian2019,airapetian2013a,airapetian2013m}. We note that
these physics techniques simultaneously account for non-uniform
(including incomplete) sampling as well as measurement bias, i.e.,
where the difference between true and measured values are not
negligible.

One challenge to bringing the histogram based methods to biomedical research is that
the techniques require explicit measurement of all confounding factors
\citep{airapetian2013a,hoecker1996}. Unlike physics, where the total set of relevant
variables is known from basic principles, in biomedical
research, the full set of relevant confounding variables is generally
not known and this assumption fails.

The goal of this manuscript is to present a method to mitigate the effect of
incomplete or otherwise non-uniform sampling when computing circular
statistic moments and testing their statistical significance.  We first describe how
these goals fall within the more general situation of correcting for
measurement bias (including non-uniform sampling), and we then present
the general method.  Our solution is thus applicable not just to circular statistic moments and their statistical significance but also to
any situation where the goal is to determine linear parametrizations of the true underlying distribution in the presence of non-negligible
measurement bias, including non-uniform sampling. The only requirement is the ability to simulate the
measurement bias.

\section{Methods}

\subsection{Overview}

\begin{figure}
  \begin{center}
  \includegraphics[width=0.9\textwidth]{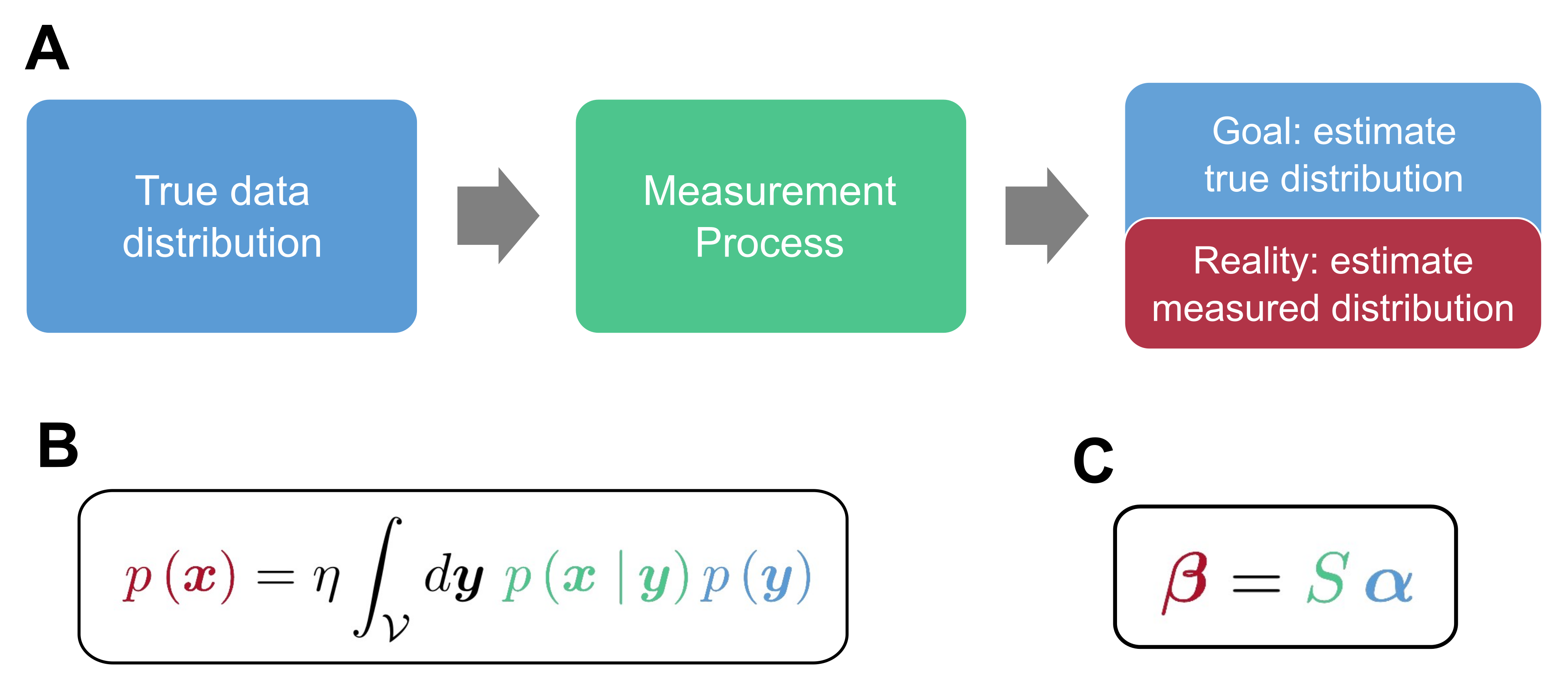}
  \caption{{\bf Overview of the general problem and solution.} A) Graphical description of the measurement process.  We focus on situations where there is an underlying true data distribution which we would like to estimate.  We conduct some measurements with the goal of assessing properties of that true distribution.  However, what we directly assess is the distribution of the measured values, biased to some degree by the measurement process.  Thus, we need a method to mitigate this bias.  Color coding is as follows: blue, aspects related to the true distribution; red, the reality of what is measured; green, the measurement process, which is the connection between the true distribution and what is measured.
  B) Mathematical representation.  The underlying true distribution is written as $\pxT$, where $\bm y$ is the true value of the relevant variables. The measurement process is modeled as a conditional probability $\pxMT$, the probability that value $x$ is measured when the true value was $\bm y$.  The actual measured distribution, $\pxM$, can be written as an integral of these other two quantities.  C) Linear equation.  By choosing appropriate parametrizations of the distributions in (B), the integral can be reduced to a linear equation. The measured experimental data yield parameters $\bm\beta$, and simulations allow us to estimate the effect of measurement, denoted the smearing matrix $S$.  This allows us to invert the equation to solve for $\bm\alpha$, the parameters of the true underlying distribution.  }
  \label{fig:overview}
  \end{center}
\end{figure}

A summary figure is presented in Fig. \ref{fig:overview} which provides a high-level overview of the general problem and our solution.  While we borrow heavily from techniques and concepts in the field of high energy physics, our unique contribution is mapping these ideas back to underlying integral equations and using basis functions rather than histograms to estimate these integrals. The remainder of the methods section contains the full mathematical derivation of our solution.  Note, rather than making new terms for various mathematical entities and technical concepts, we use several terms common in the high energy physics community. As these terms may be unfamiliar to many readers, we provide a brief glossary in Table \ref{tab:glossary}.

\begin{table}
    \centering
    \begin{tabular}{p{0.25\textwidth}|p{0.65\textwidth}}
       {\bf Term}  & {\bf Definition} \\\hline\hline
        Unfolding & The process of mitigating the effect of non-uniform sampling. \\\hline
        Experimental, measured, or uncorrected data & These names all refer to the data directly obtained from the measurement process, i.e., the data directly prior to unfolding. \\\hline
        Acceptance & A term referring to the sampling process: perfect acceptance means to uniform (unbiased) sampling, limited acceptance means incomplete sampling, and non-uniform acceptance means non-uniform sampling.  \\\hline
        Smearing Matrix & The matrix representing the measurement process, a parametrization of the conditional probability $\pxMT$.\\\hline
        Cross-talk & Correlation between otherwise independent extracted parameters induced by the non-uniform sampling.
    \end{tabular}
    \caption{{\bf Glossary of terms}}
    \label{tab:glossary}
\end{table}

\subsection{Background Information}
\label{sec:background}

Before introducing our new methods, we first review Monte Carlo integration and some details about circular moments. This allows us to highlight the close relationship between circular moments, Fourier moments, and the Rayleigh test statistic.

Direct Monte Carlo integration is a common technique for numeric estimation of integrals
\citep{cappe2005}. Consider an integral of the form
\begin{equation}
  \mathcal I = \int d{\bm x}\ p(\bm x) g(\bm x),
\end{equation}
where the integration domain is a
real, multidimensional space, and $p(\bm x)$ is a probability distribution function.  Based on the law of the
unconscious statistician (LOTUS), the integral is simply the
expectation value of $g(x)$ given $p(x)$, i.e., $\mathbb E_p\left[g(\bm x)\right]$. Given $n$
data points $\left\{{\bm x}^{(k)}\right\}_{k=1}^n$ drawn from $p(\bm
x)$, the numeric estimate of the integral is
\begin{equation}
  \label{eq:Ihat}
  \hat{\mathcal I} = \frac{1}{n} \sum_{k=1}^n g\left({\bm x}^{(k)}\right),
\end{equation}
\citep{cappe2005}.

The definition of the $k$th circular moment is 
\begin{eqnarray}
  \label{eq:c_expectation}
  c_k &=& {\mathbb E}_{p(\phi)}\left[e^{-i k\phi}\right],\\
  \label{eq:c_integral}
      &=& \int_{-\pi}^{\pi} d\phi\ p(\phi)\, e^{-i k\phi},
\end{eqnarray}
where $p(\phi)$ is a one-dimensional circular probability distribution function (PDF) with domain spanning $(-\pi,\pi]$. Note, Eq. \ref{eq:c_integral} follows from Eq. \ref{eq:c_expectation} due to LOTUS.  We also observe from Eq. \ref{eq:c_integral} that the circular moments are identical to the Fourier moments of the PDF $p(\phi)$.  Thus, the PDF can be formally written as
\begin{equation}
p(\phi) = \frac{1}{2\pi} \sum_{k=-\infty}^\infty c_k\, e^{i k\phi}.
\end{equation}
Note, normalization of the PDF implies that $c_0 = 1$.  Lastly, as $\phi$ is assumed to be real-valued, the $k$ and $-k$ components are equal, and thus we arrive at
\begin{equation}
   \label{eq:p_model_1}
   p(\phi) = \frac{1}{2\pi} \left( 1 +  2 \sum_{k=1}^\infty c_k\, e^{i k\phi} \right).
\end{equation}

We can apply direct Monte Carlo integration to Eq. \ref{eq:c_integral}
to obtain the standard formula for the numeric estimate of the
circular moments.  Let $\{\phi_j\}_{j=1}^n$ be a sample of $n$
data points drawn from $p(\phi)$.  Applying direct Monte Carlo
integration yields the following formula as an estimate of $\hat{c}_k$, the
numerical estimate of the true coefficients $c_k$:
\begin{equation}
  \label{eq:c_sum}
  \hat{c}_k = \frac{1}{n} \sum_{j=1}^n e^{-i k \phi_j}.
\end{equation}
We note, in practice, it easier to work with the real valued components
\begin{eqnarray}
  \label{eq:a_raw_moments}
\hat{a}_k &=& \frac{1}{n} \sum_{j=1}^n \cos(k \phi_j), \\
  \label{eq:b_raw_moments}
\hat{b}_k &=& \frac{1}{n} \sum_{j=1}^n \sin(k \phi_j),
\end{eqnarray}
with $\hat{c}_k = \hat{a}_k + i\hat{b}_k$.  Note, if we were to model the PDF as
\begin{equation}
  \label{eq:p_model_2}
  p(\phi) = \frac{1}{2\pi}\left[ 1 + \sum_{k=1}^\infty a_k \cos(k\phi) + b_k \sin(k\phi) \right],
\end{equation}
we then have the relationships
\begin{eqnarray}
  \label{eq:a_ahat}
  a_k &=& \hat{a}_k + \hat{a}_{-k} = 2\hat{a}_k,\\ 
  \label{eq:b_bhat}
  b_k &=& \hat{b}_k + \hat{b}_{-k} = 2\hat{b}_k;
\end{eqnarray}
note the factor of 2 consistent with Eq. \ref{eq:p_model_1}.
Lastly, the Rayleigh test statistic, often written $n R^2$, is defined as the number of data points times the magnitude of the first moment \citep{brazier1994},
\begin{equation}
  \label{eq:Rayleigh}
  n R^2 = n\,|c_1|^2 = n\left(\hat{a}_1^2+\hat{b}_1^2\right).
\end{equation}
The statistic $2 n R^2$ follows a $\chi^2$-distribution with 2 degrees
of freedom \citep{brazier1994}.

Thus, we see that the standard formula for circular statistics are
just estimates of the Fourier moments using the method of direct Monte
Carlo integration.  The Raleigh test statistic directly follows from
the first moment.  In the case of complete sampling and non-negligible
measurement bias, the standard method is to estimate the Fourier
moments using Eqs. \ref{eq:a_raw_moments}-\ref{eq:b_raw_moments}, from
which the Rayleigh test statistic can be derived.  In the case of
incomplete or otherwise non-uniform sampling or non-negligible
measurement bias, these Fourier moments cannot be estimated directly
from Eqs. \ref{eq:a_raw_moments}-\ref{eq:b_raw_moments}; instead, more
advanced techniques are needed---the gap directly addressed by this
manuscript.  We note that our proposed correction method is more
general than just computing Fourier moments, and so we next present
this more general case.  The specific case of Fourier moments is
considered afterwards in Section \ref{sec:Fourier_Basis}.

\subsection{The Unfolding Procedure}
\subsubsection{Theoretical Foundations}

We are now ready to present the theoretical basis of our correction
method.  Consider a general experiment based on measuring the
occurrence of events.  For each event, let certain qualities be
measured, represented by the vector $\xM$, for example, the time of
day the event occurred.  Consider also the inaccessible vector of true
values $\xT$. In other words, if a perfect measurement process and the
actual measurement process both measured the exact same circumstance,
the perfect measurement process would yield $\xT$ and the actual
measurement process would either yield $\xM$ or fail to record the event.  The difference between $\xT$
and $\xM$ reflects the precision and accuracy of the measurement
process.  While one can directly estimate the PDF of the measured
values, $\pxM$, the goal is to estimate the PDF of the true values,
$\pxT$, i.e., the values that would be obtained given a perfect
experiment.  The relationship between the true and measured PDFs can
be expressed as a Fredholm integral equation,
\begin{equation}
  \label{eq:Fredholm}
  \pxM = \eta \int_{\mathcal V} d\xT\ \pxMT\pxT.
\end{equation}
where $\mathcal V$ is the measurement domain with volume $V$. The
conditional probability $\pxMT$ captures both measurement bias and
non-uniform (including incomplete) sampling. The factor of $\eta$ is present as
an overall scale factor to ensure the PDF on the left is properly
normalized in the case of non-uniform sampling. In the Fredholm
equation, Eq. \ref{eq:Fredholm}, we consider that $\pxM$ and $\pxMT$ are 
known quantities and that $\eta$ and $\pxT$ are unknown. We note Fredholm
integral equations are inherently ill-conditioned
\citep{polyanin2008}.  In practice, however, a sufficiently accurate
approximation can often be found, but this accuracy must always be
checked; see Sec. \ref{sec:design}.

We next introduce a generic set of $N$ basis functions
$\{f_i(\bm x)\}_{i=1}^N$.  We note $\{f_i(\bm x)\}_{i=1}^N$ need not be a
complete basis of the entire space of PDF functions (i.e., the space
of $L_1$ integrable functions), nor is it required that the basis be
orthonormal.  It is sufficient that it spans a subset covering the
expected domain of $\pxM$ and $\pxT$ and that the inner product matrix $F_{i,j} = \int\ d\bm x\ f_i(\bm x)f_j(\bm x)$ be full rank.

Our approach determines first an estimate of $\eta\, \pxT$, from which the estimates of $\eta$ and $\pxT$ can be derived by noting
\begin{equation}
  \eta = \int_{\mathcal V} d\xT\ \eta\, \pxT.
\end{equation}
Let $\eta\, \pxT$ be parameterized such that
\begin{equation}
  \label{eq:eta_pxT}
  \eta\, \pxT = \sum_{i=1}^N \alpha_i\, f_i(\xT).
\end{equation}
Next, let matrix $S$ and vector $\bm\beta$ be defined such that
\begin{eqnarray}
  \label{eq:beta}
  \beta_i &=& \int d\xM\ \pxM f_i(\xM) \\
  \label{eq:S}
  S_{i,j} &=& \int d\xM\, d\xT\ \pxMT f_i(\xM) f_j(\xT).
\end{eqnarray}
This matrix $S$ is a generalization of the smearing matrix of \citet{airapetian2013a}.
We can then multiple both sides of the Fredholm integral equation with
a basis function $f_k(\xM)$ and integrate over $\xM$, which simplifies
to
\begin{equation}
  \label{eq:linear_eq}
  \bm \beta = S\, \bm \alpha.
\end{equation}
Solving for $\bm\alpha$ is then simple linear algebra, with various applicable techniques discussed in the Section \ref{sec:solution}.  First, however, we describe estimation of $S$ and $\bm\beta$ and their covariance.

%
%

\subsubsection{Estimation of the known quantities in the Fredholm Equation}
The quantities considered as known in the Fredholm Equation
(Eq. \ref{eq:Fredholm}) are $\pxM$ and $\pxMT$, with parameters in the vector $\bm\beta$ and the matrix $S$.
We next estimate this matrix and vector using direct
Monte Carlo integration. Let $\{\xM^{(k)}\}_{k=1}^n$ be a set of $n$
data points drawn from the distribution $\pxM$.  Note, these are the values measured in the experiment.
Elements of the vector $\bm\beta$ are estimated according to
\begin{equation}
  \label{eq:beta_estimate}
  \beta_i = \frac{1}{n} \sum_{k=1}^n f_i\left(\xM^{(k)}\right).
\end{equation}
We note that $p(\xM,\xT)$ can be expressed as $p(\xM|\xT) = p(\xM,\xT) / p(\xT)$.  Thus, using a numeric simulation, one can generate data
according to $p(\xM,\xT)$ with a uniform prior, i.e., $p(\xT) = 1/V$, where $V$ is the volume of the integration domain.  In
practice, this typically means starting with a data set of $\xT$ drawn
uniformly at random over the full measurement domain $\mathcal V$, redacting
data points to mimic the effect of incomplete or otherwise non-uniform sampling, and
estimating the measured value $\xM$ for each $\xT$ that was not redacted.

Let $\{\xM^{\prime (k)},\xT^{\prime (k)}\}_{k=1}^m$ be a set of $m$ data points drawn from $p(\xM,\xT)$ with uniform $p(\xT)$.  Elements of the
matrix $S$ can be then computed according to
\begin{equation}
  \label{eq:S_estimate}
  S_{i,j} = \frac{V}{m} \sum_{k=1}^m f_i\left(\xM^{\prime (k)}\right) f_j\left(\xT^{\prime (k)}\right).
\end{equation}

\subsubsection{Solving the Fredholm Equation (Unfolding)}
The process of solving the Fredholm Equation (Eq. \ref{eq:Fredholm})
or numeric approximates (Eq. \ref{eq:linear_eq}) is often called
\emph{unfolding} in the high energy physics community (see Table \ref{tab:glossary}).
\label{sec:solution}
Based on Eq. \ref{eq:linear_eq}, the solution for $\bm \alpha$ can formally be written as
\begin{equation}
  \label{eq:alpha_formal}
  \bm \alpha = S^{-1} \bm \beta.
\end{equation}
In practice, a more stable estimate of alpha can be determined using
methods to solve the linear equation that do not involve the inverse
of $S$ (such as QR decomposition).  We note that if the null-space of
$S$ is non-trivial, then $\bm\alpha$ is not unique. In practice, $S$ tends to be full rank but can have
eigenvalues near-zero.  The condition number of the matrix $S$ (ratio
of the largest to smallest eigenvalue) thus directly assesses whether
the bias and non-uniform sampling preclude a unique estimation of
$\bm\alpha$, i.e., $\pxT$.

\subsubsection{Fourier Basis}
\label{sec:Fourier_Basis}
To return to the case of circular statistics and the Rayleigh test, we
consider the case of the basis functions $\{f_k\left(\xM\right)\}$
being Fourier moments. Let $N^\prime$ be the maximum order, and we enumerate the $N=2N^\prime+1$ basis
functions as
\begin{eqnarray}
  f_k\left(\xM\right) &=& \left\{
    \begin{array}{ll}
      \cos\left(k\xM\right),            & 0 \le k \le N^\prime \\
      \sin\left((k-N^\prime)\xM\right), & N^\prime+1 \le k \le N.
    \end{array}
    \right.
\end{eqnarray}
When using this basis for unfolding, Eq. \ref{eq:eta_pxT} implies that
$\eta = \alpha_0$.  The remaining elements of $\alpha$ are proportional to $a$ and $b$ from
Eq. \ref{eq:p_model_2}.  To recover the circular statistics, we just
need to divide $\bm\alpha$ by a factor of $2\alpha_0$ and focus on indices greater than zero,
\begin{equation}
  \label{eq:alphaprime}
  \alpha_i^\prime = \frac{\alpha_i}{2 \alpha_0 }\ \forall \ i=[1,2,\dots N].
\end{equation}
We can also transform the circular moments $\bm\alpha^\prime$ into magnitudes and phases, e.g.,
\begin{eqnarray}
  \label{eq:mag}
  |c_k|  &=& \sqrt{ \alpha^{\prime\, 2}_k + \alpha^{\prime\, 2}_{k+N^\prime}} = \frac{1}{2} \sqrt{ \frac{\alpha^2_k+\alpha^2_{k+N^\prime}}{\alpha_0^2} },\\
  \varphi_k &=& \tan^{-1}\left( \frac{\alpha^\prime_{k+N^\prime}}{\alpha^\prime_k} \right) = \tan^{-1}\left( \frac{\alpha_{k+N^\prime}}{\alpha_k} \right).
\end{eqnarray}
Note, the Fourier moments are $2|c_k|$ (see Section \ref{sec:background}).

Lastly, the Raleigh test statistic can be computed by substituting the values from Eq. \ref{eq:mag} into Eq. \ref{eq:Rayleigh}. Note, however, that the Rayleigh test statistic does not account for the measurement effects
captured in $\pxMT$, and thus it may overestimate the significance (i.e., underestimate the $p$-value).  See more details in our fourth simulation, described in Section \ref{sec:sim4}.

\subsubsection{Design Considerations}
\label{sec:design}
Given the inherent ill-conditioned nature of the incomplete sampling
and measurement bias captured in $\pxMT$ and the matrix $S$, it is
essential that every analysis using this method assess the stability
and accuracy of the solution given their unique measurement scenario.
This involves two specific tasks.  First, the condition number of the
matrix $S$ must be considered.  Secondly, simulated data should be
generated with a known true distribution, and then the moments from
the known distribution should be compared with the reconstructed
values. The method should be repeated for various values of the true
distribution as more than one true distribution could result in the
same measured distribution if the null space of $S$ is approximately
non-trivial.  In these cases, it is wise to generate a large amount of
data---10 times the among of actual experimental data is a common rule
of thumb---such that the variance on the estimated parameters from the
simulation is much lower than the variance of the parameters extracted
from the actual experimental data. Specific examples of conducting
these simulations are given in Section \ref{sec:exp}.

A potentially overlooked aspect of non-uniform
or incomplete sampling is that it can induce correlation in measured
variables even when the true variables are not correlated. This effect
is sometimes called cross-talk. This should be remembered when
reporting results, as all extracted coefficients are correlated. This
correlation also impacts the selection of the basis functions.  For
example, if one desires to assess circadian cyclicity, it is sufficient to
estimate the true value of the magnitude of the first Fourier
moment. One might naively select a basis which only includes the first
moment, i.e. a basis of $\sin(\phi)$ and $\cos(\phi)$.  However,
giving the mixing of moments caused by the nontrivial $\pxMT$, it is
necessary in practice to include the zeroth moment (which should
always be included) as well as potentially as higher moments.
Conceptually, this is similar to aliasing, as the higher moments
influence the estimate of the lower moments, although the cause is
completely different. Thus, there is a trade-off in the choice of
basis functions: too few, and the effect of the mixing between moments
will not get accurately unfolded; but too many, and the condition
number of $S$ will increase.  A good design practice is to select
enough basis functions to allow the coefficients (moments) of interest
to be reconstructed with sufficient accuracy, but not to include so
many as to significantly negatively impact the condition number of the smearing matrix
$S$.  The optimal number of basis functions can potentially vary for
each data set considered.

\section{Code availability}
Matlab code has been posted to \url{https://github.com/sgliske/unfolding} .

\section{Experiments and analyses}

Each of the following simulations are based on the scenario of measuring the circadian oscillation of the occurrence of some event.  For each simulation, we repeated the same general process, which involved the following steps.  First, we chose an acceptance to use for the simulation.  We note that in actual experiments, the acceptance is a known quantity.  We then selected a known true distribution and simulated how data with that true distribution would appear if measured with the chosen acceptance. We also applied the unfolding procedure and compared how well the unfolded parameters matched the true parameters.  The process was then repeated for a variety of parameters of the true distribution and/or a variety of choices of acceptance.

\label{sec:exp}
\subsection{Simulation 1, First Toy Model}

The first simulation is a very simple toy example. For the acceptance, we simulated that the machine stopped recording during the last 6 hours of recording, and thus only the data during the first 18 hours were actually measured. In other words, the acceptance was modeled as perfect from 0.00 to 18:00 hours and exactly zero from 18:00 to 24:00 hours.  We then considered how several different true distributions would appear with this acceptance and how well the unfolding procedure would work. We specifically used an amplitude of 0.3 and selected 12 values for the zenith of the true data distribution, ranging from 1:00 to 23:00 in two hour steps.  We also used 100,000 events for estimating the smearing matrix and simulated 100,000 ``measured'' events.  We selected large numbers to minimize the impact of random fluctuations on these examples.  

\begin{figure}
  \begin{center}
  \includegraphics[width=0.95\textwidth]{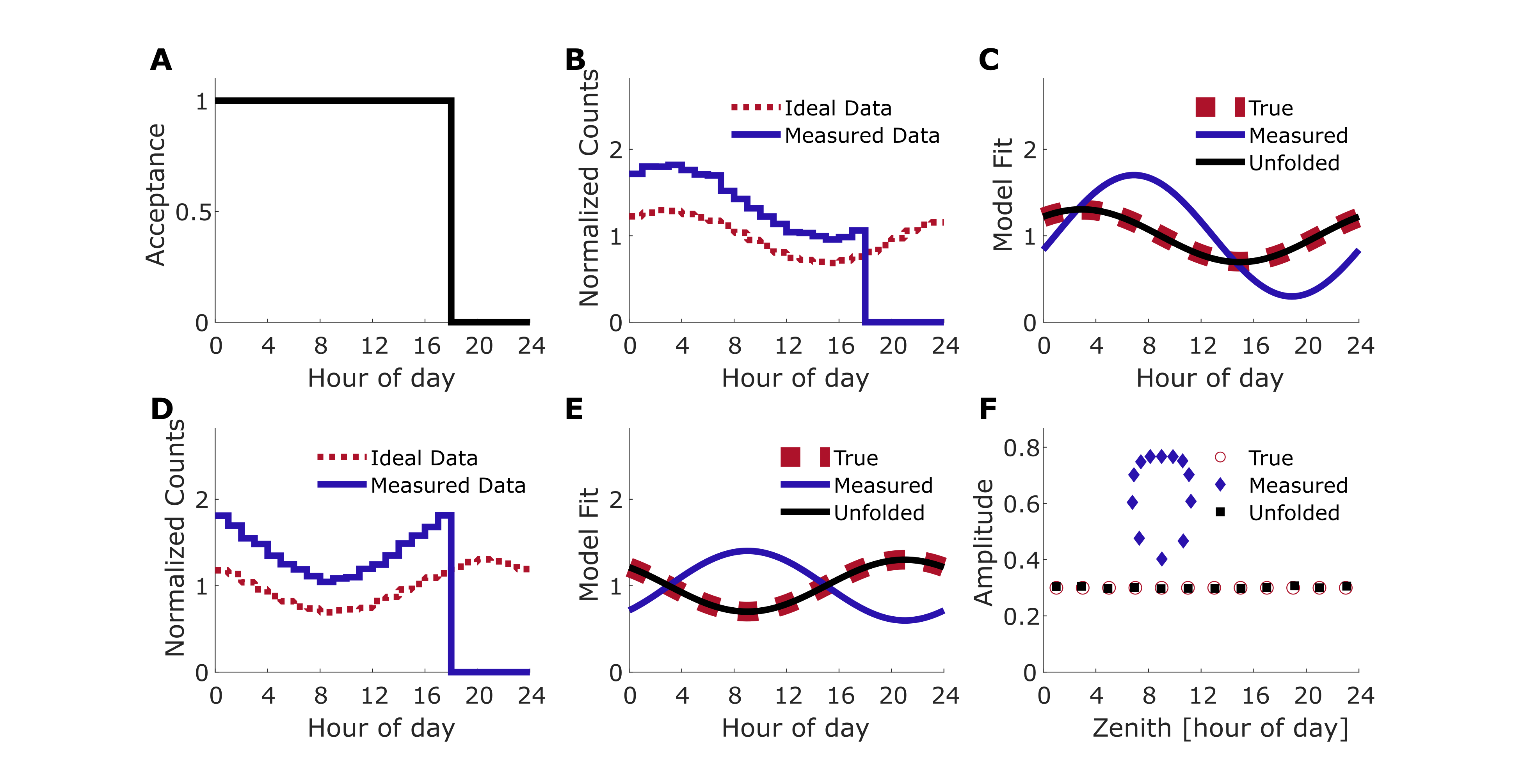}
  \caption{{\bf Results for Simulation 1, First Toy Model.}  A) Acceptance.  In this simulation, data were recorded uniformly from 0.00 to 18:00 hours, but no data was recorded from 18:00 to 24:00 hours. B) Data distribution, example 1. Red dotted line represents what would have been measured with ideal acceptance, whereas the blue solid line represents what would have been recorded with the acceptance in (A). Data were generated with an amplitude of 0.3 and a zenith at 3:00 hours.  C) Model fits of the data in (B), c.f., Eq. \ref{eq:p_model_2}.  While the measured data (blue line) does not match the true model (red dotted line), the model using the unfolded parameters matches extremely well. D) Data distribution, example 2. The nadir of the distribution was set to 21:00, centered in the period with no acceptance.  E) The effect is a measured data distribution that is perfectly out of phase, as the data are now symmetric about 9.00 and 21:00, and there is more data around 9.00 than around 21:00.  F) Results of the full scan over the zenith of the true distribution. We repeated the process for 12 different values of the zenith of the true distribution, spaced every two hours across the 24 hour period. The data from (C) and (E) each generate a triplet of true, measured and unfolded points in this figure, with letters next to the specific points corresponding to those panels.  The very large discrepancy between measured values (blue diamonds) and true values (red circles) shows that the non-uniform acceptance causes extreme bias to the measured circadian moments. }
  \label{fig:sim1}
  \end{center}
\end{figure}

Results are shown in Fig. \ref{fig:sim1}.  Even though we tested with true distributions having zenith's across the full 24 hour period, the measured moments always had a zenith between 6:00 and 12:00, consistent with a nadir always being between 18:00 to 24:00, the period when no data was recorded.  The unfolding method recovers the true zenith and amplitude.  Over all parameters considered, we observed that root mean square (RMS) residual between the true and unfolded moments was 0.004, corresponding to a 1\% resolution (0.0004/0.3 = 0.013).

\subsection{Simulation 2, Second Toy Model}

\begin{figure}
  \begin{center}
  \includegraphics[width=0.95\textwidth]{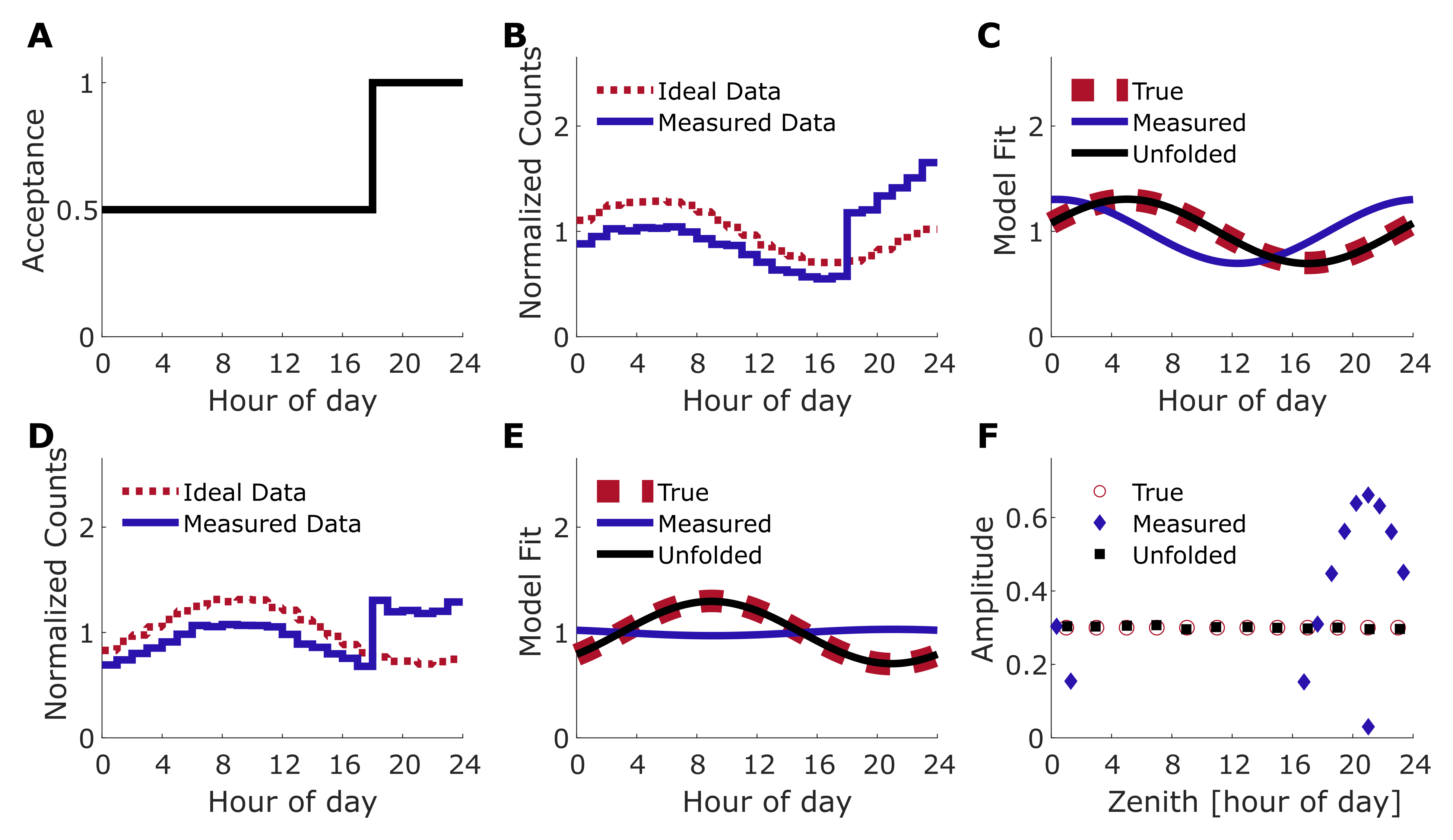}
  \caption{{\bf Results for Simulation 2, Second Toy Model.} Data are presented in the same way as they were for simulation 1, Fig. \ref{fig:sim1}. A) Acceptance. B\&D) Data distributions for two examples. C\&E) model fits for two examples.  F) Results for the full scan. The measured data always yield zeniths within a few hours of 21:00 with widely varying amplitudes, even though the true zeniths span the full 24 hour period and the true amplitudes were always 0.3. In the first example (B-C), the amplitude is nearly unchanged by the zenith is shifted from 5:00 to 0:24 hours. In the second example (D-E), the heightened acceptance occurs right at the zenith of the distribution, causing the amplitude of the measured data to be near zero and the zenith to be off by 12 hours. Unfolding is able to reconstruct the true values with small residual.}
  \label{fig:sim2}
  \end{center}
\end{figure}

In the second simulation, we repeated the identical procedure as in the first simulation but with a different choice of acceptance.  We modeled that the recording device was started at 18:00 hours, and turned off at 24:00 hours, a recording duration of 30 hours. The acceptance is thus twice as high during 18:00 to 24:00 hours as it is during the rest of the day. See Fig. \ref{fig:sim2}.  We observed that depending on the zenith of the true distribution, the acceptance can cause the measured data to have circular moments with amplitudes nearly unchanged (but with a shifted zenith), Fig. \ref{fig:sim2}B-C, or with amplitudes near zero and a zenith that is off by 12 hours, Fig. \ref{fig:sim2}D-E. As in simulation 1, the unfolding procedure recovers the correct parameters with a very lower error: the residual RMS is again 0.004. 

\subsection{Simulation 3: Real World Acceptance}
\label{sec:sim3}

\subsubsection{Motivation}
For our next simulation, we will use acceptance values from real world data related to our ongoing epilepsy research. High frequency oscillations (HFOs) are an
electrographic element observed in intracranial EEG
\citep{worrell2004}.  While healthy tissue can produce HFOs, they have a
higher prevalence in tissue instigating seizures
\citep{engel2009,gliske2016a,gliske2020}. Understanding how HFO rates vary with time of day
and with state of vigilance gives insight into their pathophysiology
and may also help guide optimization of their clinical
interpretation. However, intracranial EEG monitoring typically involves gaps in
the recordings, for example, if the patient needs to have mapping
procedures or extra imaging.  Thus, the sampling across the 24-hour
period is not uniform.  Additionally, patients sleep very poorly
during intracranial EEG monitoring, and thus restricting HFO analysis
to specific stages of sleep results in highly inconsistent and
sporadic sampling of events across the 24-hour daily cycle; see Fig. \ref{fig:example}.  Thus,
methods to correct for non-uniform and incomplete sampling are
essential to understand the cyclicity of HFOs controlled for state of
vigilance. 

\subsubsection{Patient data}
At the University of Michigan, we have acquired a large database of multi-day, intracranial EEG recordings, the vast majority of which have had sleep scoring
performed by a sleep technician based on the scalp EEG
\citep{gliske2018}. The
database was gathered under the approval of the local Institutional
Review Board (IRB), and all subjects in the database have either given
their consent (adults) or have assented to participate (children) with
consent being provided by a parent, guardian, or legally authorized
representative. All subjects meeting the following inclusion criteria
as of February 1, 2022 were included in this study: clinical data
acquisition with sampling rate of 4,096 Hz and sleep scoring completed
for at least 24 hours of data. This resulted in 58 subjects, with the
amount of sleep-scored data ranging from 30.1 to 395.0 hours (median
167 hours). We divided state of vigilance into 3 categories: awake, REM,
and non-REM (NREM).  Only data from one subject is used in this simulation, Simulation 3, though data from all subjects are used later in Simulation 4.

\subsubsection{Experimental design }
We repeated the process used in simulation 1 and 2 with two minor modifications.  First, for the acceptance we selected is that of when NREM sleep occurred in an example patient. Note, in actual experiments, the acceptance is a known quantity. Second, we increased our parameter scan for the true data distribution.  For each of the 12 zenith values, we also considered 5 different amplitude values, ranging from 0.1 to 0.5 in steps of 0.1.  See Fig. \ref{fig:sim3}. This type of study is essential when using the unfolding methods because it informs whether the non-uniformity of sampling allows or precludes reconstruction of the actual true values.

\subsubsection{Results}

\begin{figure}
  \begin{center}
  \includegraphics[width=0.95\textwidth]{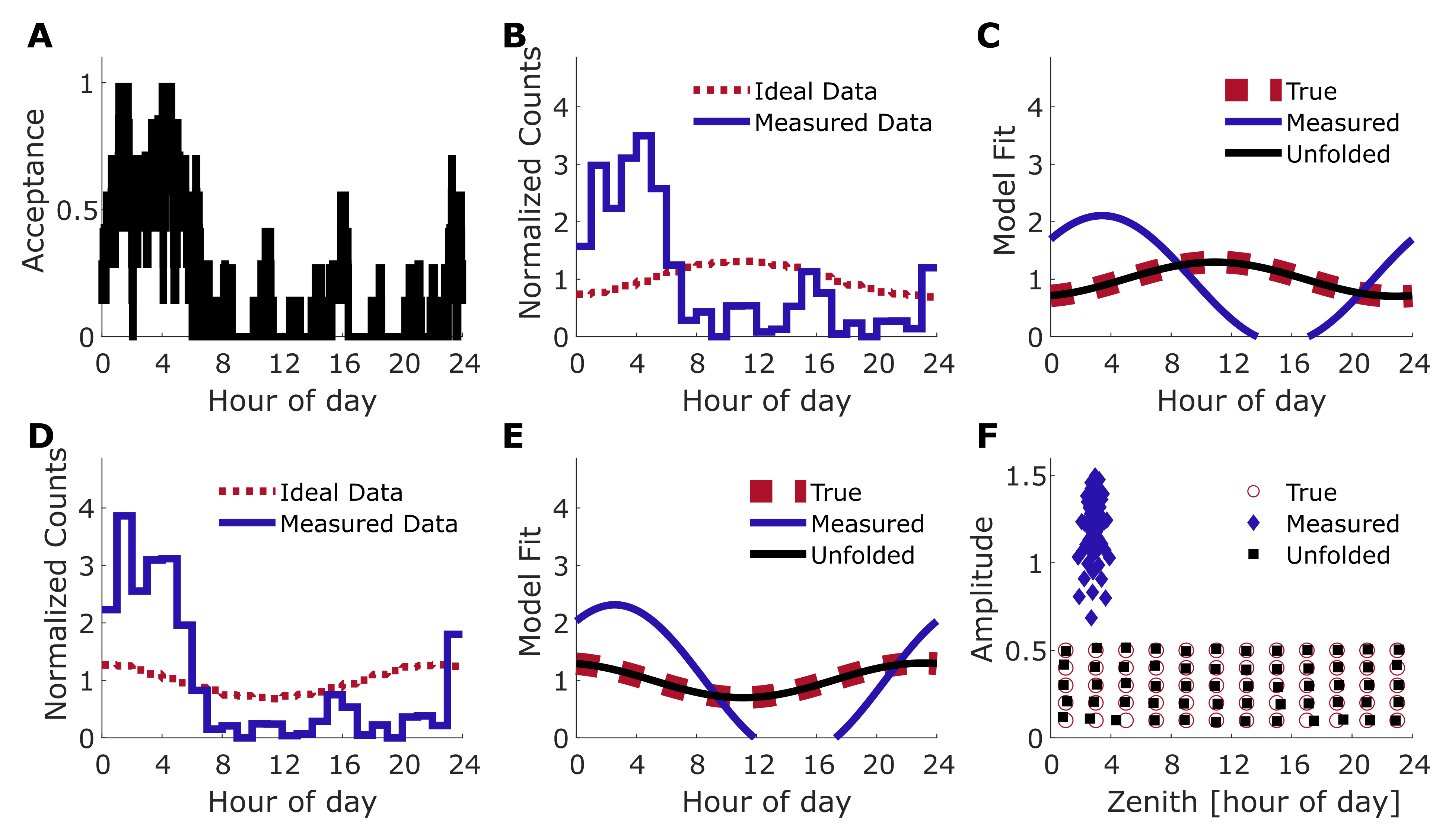}
    \caption{{\bf Results for Simulation 3, Real World Acceptance.} Data are presented in the same way as they were for simulation 1\&2, Figs. \ref{fig:sim1}-\ref{fig:sim2}. A) Acceptance. In this case, the acceptance is the relative amount of NREM sleep over the 24 hour period from a multi-day hospital stay of an example subject, the same data shown in Fig. \ref{fig:example}A.  B\&D) Data distributions for two example true distributions. C\&E) model fits for two examples.  Note, the model fits are not strictly positive, which breaks the positivity constraint for PDFs. F) Results for the full scan.  Due to the density of points, the specific values corresponding to the results in panels (C) and (E) are not indicated. We observe that the acceptance is so extreme that the measured data always yielded circular moments with zeniths between 1:00 and 5:00 hours and with amplitudes that are much too large. Unfolding is able to reconstruct the true values with small residual: RMS of residual over all true parameters is 0.007.}
  \label{fig:sim3}
  \end{center}
\end{figure}

Results for simulation 3 are shown in Fig. \ref{fig:sim3}.  The NREM sleep in this patient mainly occurred during the hours of 0:00 and 7:00 hours, but the acceptance is highly non-uniform. As with our first two simulations, we observed that circadian moments directly computed from the measured data are extremely inaccurate due to the very non-uniform acceptance.  Again, unfolding is able to reconstruct the true values with low error across the full parameter range. 

\subsection{Simulation 4: Assessing Statistical Uncertainty}
\label{sec:sim4}

\subsubsection{Motivation}
Our first several simulations have demonstrated that our unfolding procedure accurately reconstructs the circular moments.  In our last simulation, we seek to assess whether unfolding has mitigated the influence of imperfect acceptance Rayleigh test results. Recall, the Rayleigh test is an analytic approach to assess the statistical significance of circular moments being non-zero, against the null hypothesis that the moments are zero. Non-uniform acceptance increases the uncertainty, which is not accounted for in the Rayleigh test statistic. Therefore, we anticipate that the Rayleigh test statistic overestimates statistical significance and is thus is not applicable in the case of non-uniform sampling. In this simulation, we will use the Rayleigh test statistic to compute the smallest amplitude moment that is statistically significant for a given number of data points.

Numeric approaches also exist, in contrast to the analytic approach of the Rayleigh test.  Specifically, one can numerically simulate the null hypothesis for a given number of data points and for a given acceptance, repeating the process many times. For each iteration, one can unfold the moments and determine the amplitude.  The 95\%-tile of the distribution of those amplitudes gives a direct, numeric estimate of the threshold for statistical significance.  By design, this approach includes all the effects of the non-uniform acceptance for the specific acceptance used in the simulations.  In this, our fourth simulation, we compare the threshold for statistical significance computed using this numeric approach with the threshold from the Rayleigh test statistic.  

\subsubsection{Simulation design}

The simulation is designed to numerically simulate thresholds for statistical significance.  Specifically, we estimate the null hypothesis for each state of vigilance and each patient and for the number of events (HFOs) actually recorded during each of those states of vigilance in each patient. As with the other simulations, we again selected a specific acceptance and then simulated various true distributions and how they would appear with that acceptance. As with Simulation 3, we again used patient sleep data for the acceptance.  However, for Simulation 4, we repeated the process for each state of vigilance for each patient, rather than for just one example state of vigilance from one patient.  Also, rather than scanning over a variety of true distributions, for Simulation 4 we only simulated the null hypothesis (uniform true distribution).  However, we repeated the process 1,000 times for each acceptance considered (i.e., each state of vigilance for each patient).  Also, rather than using 100,000 simulated events, we now use the number of events equal to the number of actual HFOs recorded.  Thus, our results are the thresholds which will be needed in the future to assess statistical significance of unfolded moments for HFOs in these states of vigilance in these patients. 

\subsubsection{Results}

\begin{figure}
  \begin{center}
  \includegraphics[width=\textwidth]{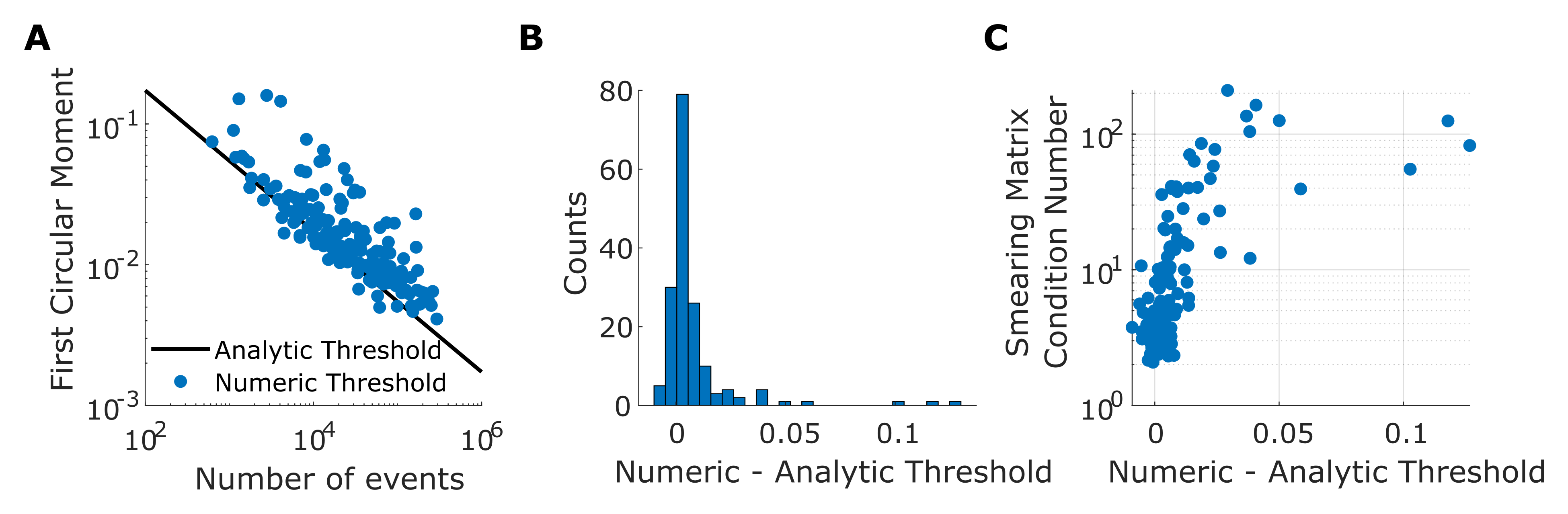}
  \caption{{\bf Thresholds for distinguishing from the
      null-hypothesis.} A) The thresholds for the amplitudes of the circular moments being statistically significant at the $\alpha=0.05$ level.  The analytic threshold, based on the Rayleigh test Statistic, appears as a straight line (black line) when plotted on a log-log plot.  The numeric threshold (blue circles), computed by simulating the null hypothesis, are typically larger than the analytic threshold.  B) Histogram of the difference between the numeric and analytic thresholds.  We observe 95\% of the differences are less than 0.032. C) Association between the difference in thresholds and the smearing matrix condition number. The Spearman correlation is
    0.84, with $p$-values below machine precision. Thus, the Rayleigh test statistic does not correctly estimate the significance, with the test statistic being farther off the more the acceptance is limited.}
  \label{fig:sim4}
  \end{center}
\end{figure}

In Fig. \ref{fig:sim4}, we compare the threshold for significance at the $\alpha=0.05$ level for the Rayleigh test statistic (``analytic method'') versus the simulation (``numeric method''). As expected, we observed that the thresholds computed by the simulation either closet to or slightly higher than those from the Raleigh Test.  At first glance, the magnitude of the difference appears small, but as the circadian moments of the HFOs have not yet been measured, one cannot conclude whether the magnitude of the difference is relatively small or not.  The difference between the thresholds from the analytic and numeric methods are strongly correlated with the condition number of the smearing matrix ($\rho=0.84$, $p$ less than machine precision, Spearman Correlation Coefficient).  Recall the condition number of the smearing matrix $S$ is an assessment of how strongly the non-uniform acceptance is limiting reconstruction of the moments. 

\section{Discussion}
We have presented a novel method to correct for non-uniform
(including incomplete) sampling and measurement bias.  We have also
shown how these methods apply to the estimation of circular
statistical moments and tests of their significance. We have applied these methods to simulations involving acceptances from simple toy models and real world data. We find that without unfolding, non-uniformity in the measurement process can have drastic influence on the
estimation of circular moments. Even so, our unfolding methods
successfully corrected for incomplete and non-uniform sampling with high
accuracy even in fairly extreme cases of incomplete and non-uniform sampling.

We observed that unfolding is not sufficient to correct the Rayleigh test statistic in all cases.  Instead, it is necessary to simulate the null hypothesis to determine statistical significance of unfolded moments.  The Rayleigh test should thus not be used with unfolded moments. 

In some cases, it is more relevant to know the confidence interval around the unfolded moments rather than whether the null hypothesis can be rejected.  This is particularly important when comparing different experiments to see if the circadian moments are consistent across the experiments. An alternate method can be used to estimate these confidence intervals, which is provided in the Supplement.

An important component of the method is selecting an
appropriate basis expansion, including the number of basis elements.
In the case of circular statistics, the basis functions are sine and
cosine functions, but which functions are most relevant should be
considered for each use case. For example, some histogram methods used
in high energy physics (e.g., \citet{airapetian2013a}), can be
represented as a special case of our method using a basis of piecewise
constant functions, though spherical harmonics are more appropriate in
some circumstances (e.g., \citet{gliske2013p}).  With any choice of
basis function, the condition number of the smearing matrix is an
essential key to selecting the number of basis elements.

One challenge that our method does not directly overcome is
related to the concept of aliasing and is ubiquitous to analyses with
incomplete sampling. In all cases, one should
simulate data with all the moments which are expected to be non-zero
to ensure that cross-talk between moments is not unduly influencing
the results.  Thus, while no method can fully replace what is lost by
incomplete sampling, careful use of simulated data allows a rigorous
assessment of when the results are expected to be reliable, thus
providing high confidence when appropriate.

One advantage of our unfolding method is that it also
allows for directly modeling how the circular static moments vary with
other factors.  One simply needs to select an appropriate set of basis
functions to model the desired dependence, and then the coefficients
for such models are all correctly adjusted at the same time.

The case of unmeasured confounding factors is worth noting.  Histogram methods are generally not valid in this case \citep{airapetian2013a,hoecker1996}.
However, the situation is the same for either unfolding with our method or directly computing circular statistics. In both cases, the moments represent the expectation value given the distribution of that unmeasured confounding factor in our data.  Thus, unmeasured confounding factors are not a limiting factor.

\section{Conflict of interest statement}

The author has licensed intellectual property related to detection of
high frequency oscillations to Natus Neurology, which has only
incidental relationship to this work.

This work was funded by the National Institute of Health (NIH), award
R01-NS094399. The human database was also developed under this as
award as well NIH awards K01-ES026839, K08-NS069783, and UL1-TR000433
and the Doris Duke Foundation award 2015096.

\bibliography{ms}

\begin{thebibliography}{16}
\expandafter\ifx\csname natexlab\endcsname\relax\def\natexlab#1{#1}\fi
\providecommand{\url}[1]{\texttt{#1}}
\providecommand{\href}[2]{#2}
\providecommand{\path}[1]{#1}
\providecommand{\DOIprefix}{doi:}
\providecommand{\ArXivprefix}{arXiv:}
\providecommand{\URLprefix}{URL: }
\providecommand{\Pubmedprefix}{pmid:}
\providecommand{\doi}[1]{\href{http://dx.doi.org/#1}{\path{#1}}}
\providecommand{\Pubmed}[1]{\href{pmid:#1}{\path{#1}}}
\providecommand{\bibinfo}[2]{#2}
\ifx\xfnm\undefined \def\xfnm[#1]{\unskip,\space#1}\fi
\bibitem[{Airapetian et~al.(2013{\natexlab{a}})Airapetian, Akopov, Akopov,
  Aschenauer, Augustyniak, Avakian, Avetissian, Avetisyan, Belostotski and
  Blok}]{airapetian2013a}
\bibinfo{author}{Airapetian\xfnm[ A.]}, \bibinfo{author}{Akopov\xfnm[ N.]},
  \bibinfo{author}{Akopov\xfnm[ Z.]}, \bibinfo{author}{Aschenauer\xfnm[ E.]},
  \bibinfo{author}{Augustyniak\xfnm[ W.]}, \bibinfo{author}{Avakian\xfnm[ R.]},
  \bibinfo{author}{Avetissian\xfnm[ A.]}, \bibinfo{author}{Avetisyan\xfnm[
  E.]}, \bibinfo{author}{Belostotski\xfnm[ S.]}, \bibinfo{author}{Blok\xfnm[
  H.]}.
\newblock \bibinfo{title}{Azimuthal distributions of charged hadrons, pions,
  and kaons produced in deep-inelastic scattering off unpolarized protons and
  deuterons}.
\newblock \bibinfo{journal}{Physical Review D}
  \bibinfo{year}{2013}{\natexlab{a}};\bibinfo{volume}{87}(\bibinfo{number}{1}):\bibinfo{pages}{012010}.
\newblock \DOIprefix\doi{10.1103/PhysRevD.87.012010}.
\bibitem[{Airapetian et~al.(2013{\natexlab{b}})Airapetian, Akopov, Akopov,
  Aschenauer, Augustyniak, Avakian, Avetissian, Avetisyan, Belostotski and
  Blok}]{airapetian2013m}
\bibinfo{author}{Airapetian\xfnm[ A.]}, \bibinfo{author}{Akopov\xfnm[ N.]},
  \bibinfo{author}{Akopov\xfnm[ Z.]}, \bibinfo{author}{Aschenauer\xfnm[ E.]},
  \bibinfo{author}{Augustyniak\xfnm[ W.]}, \bibinfo{author}{Avakian\xfnm[ R.]},
  \bibinfo{author}{Avetissian\xfnm[ A.]}, \bibinfo{author}{Avetisyan\xfnm[
  E.]}, \bibinfo{author}{Belostotski\xfnm[ S.]}, \bibinfo{author}{Blok\xfnm[
  H.]}.
\newblock \bibinfo{title}{Multiplicities of charged pions and kaons from
  semi-inclusive deep-inelastic scattering by the proton and the deuteron}.
\newblock \bibinfo{journal}{Physical Review D}
  \bibinfo{year}{2013}{\natexlab{b}};\bibinfo{volume}{87}(\bibinfo{number}{7}):\bibinfo{pages}{074029}.
\newblock \DOIprefix\doi{10.1103/PhysRevD.87.074029}.
\bibitem[{Airapetian et~al.(2019)Airapetian, Akopov, Akopov, Aschenauer,
  Augustyniak, Avakian, Avetissian, Belostotski, Blok and
  Borissov}]{airapetian2019}
\bibinfo{author}{Airapetian\xfnm[ A.]}, \bibinfo{author}{Akopov\xfnm[ N.]},
  \bibinfo{author}{Akopov\xfnm[ Z.]}, \bibinfo{author}{Aschenauer\xfnm[ E.]},
  \bibinfo{author}{Augustyniak\xfnm[ W.]}, \bibinfo{author}{Avakian\xfnm[ R.]},
  \bibinfo{author}{Avetissian\xfnm[ A.]}, \bibinfo{author}{Belostotski\xfnm[
  S.]}, \bibinfo{author}{Blok\xfnm[ H.]}, \bibinfo{author}{Borissov\xfnm[ A.]}.
\newblock \bibinfo{title}{Longitudinal double-spin asymmetries in
  semi-inclusive deep-inelastic scattering of electrons and positrons by
  protons and deuterons}.
\newblock \bibinfo{journal}{Physical Review D}
  \bibinfo{year}{2019};\bibinfo{volume}{99}(\bibinfo{number}{11}):\bibinfo{pages}{112001}.
\newblock \DOIprefix\doi{10.1103/PhysRevD.99.112001}.
\bibitem[{Baud et~al.(2018)Baud, Kleen, Mirro, Andrechak, King-Stephens, Chang
  and Rao}]{baud2018}
\bibinfo{author}{Baud\xfnm[ M.O.]}, \bibinfo{author}{Kleen\xfnm[ J.K.]},
  \bibinfo{author}{Mirro\xfnm[ E.A.]}, \bibinfo{author}{Andrechak\xfnm[ J.C.]},
  \bibinfo{author}{King-Stephens\xfnm[ D.]}, \bibinfo{author}{Chang\xfnm[
  E.F.]}, \bibinfo{author}{Rao\xfnm[ V.R.]}.
\newblock \bibinfo{title}{Multi-day rhythms modulate seizure risk in epilepsy}.
\newblock \bibinfo{journal}{Nature Communications}
  \bibinfo{year}{2018};\bibinfo{volume}{9}(\bibinfo{number}{1}):\bibinfo{pages}{1--10}.
\newblock \DOIprefix\doi{10.1038/s41467-017-02577-y}.
\bibitem[{Brazier(1994)}]{brazier1994}
\bibinfo{author}{Brazier\xfnm[ K.]}.
\newblock \bibinfo{title}{Confidence intervals from the rayleigh test}.
\newblock \bibinfo{journal}{Monthly Notices of the Royal Astronomical Society}
  \bibinfo{year}{1994};\bibinfo{volume}{268}(\bibinfo{number}{3}):\bibinfo{pages}{709--712}.
\newblock \DOIprefix\doi{10.1093/mnras/268.3.709}.
\bibitem[{Capp\'e et~al.(2005)Capp\'e, Moulines and Ryd\'en}]{cappe2005}
\bibinfo{author}{Capp\'e\xfnm[ O.]}, \bibinfo{author}{Moulines\xfnm[ E.]},
  \bibinfo{author}{Ryd\'en\xfnm[ T.]}.
\newblock \bibinfo{title}{Inference in Hiddon Markov Models}.
\newblock \bibinfo{address}{New York, NY}: \bibinfo{publisher}{Springer},
  \bibinfo{year}{2005}.
\bibitem[{Engel et~al.(2009)Engel, Bragin, Staba and Mody}]{engel2009}
\bibinfo{author}{Engel\xfnm[ J.]}, \bibinfo{author}{Bragin\xfnm[ A.]},
  \bibinfo{author}{Staba\xfnm[ R.]}, \bibinfo{author}{Mody\xfnm[ I.]}.
\newblock \bibinfo{title}{High-frequency oscillations: {What} is normal and
  what is not?}
\newblock \bibinfo{journal}{Epilepsia}
  \bibinfo{year}{2009};\bibinfo{volume}{50}(\bibinfo{number}{4}):\bibinfo{pages}{598--604}.
\newblock \DOIprefix\doi{10.1111/j.1528-1167.2008.01917.x}.
\bibitem[{Gliske and Pappalardo(2013)}]{gliske2013p}
\bibinfo{author}{Gliske\xfnm[ S.]}, \bibinfo{author}{Pappalardo\xfnm[ L.]}.
\newblock \bibinfo{title}{Dihadron production in semi-inclusive {DIS} from
  transversely polarized protons}.
\newblock \bibinfo{journal}{PoS}
  \bibinfo{year}{2013};\bibinfo{volume}{DIS2013}:\bibinfo{pages}{233}.
\newblock \DOIprefix\doi{10.22323/1.191.0233}.
\bibitem[{Gliske et~al.(2018)Gliske, Irwin, Chestek, Hegeman, Brinkmann,
  Sagher, Garton, Worrell and Stacey}]{gliske2018}
\bibinfo{author}{Gliske\xfnm[ S.V.]}, \bibinfo{author}{Irwin\xfnm[ Z.T.]},
  \bibinfo{author}{Chestek\xfnm[ C.]}, \bibinfo{author}{Hegeman\xfnm[ G.L.]},
  \bibinfo{author}{Brinkmann\xfnm[ B.]}, \bibinfo{author}{Sagher\xfnm[ O.]},
  \bibinfo{author}{Garton\xfnm[ H.J.L.]}, \bibinfo{author}{Worrell\xfnm[
  G.A.]}, \bibinfo{author}{Stacey\xfnm[ W.C.]}.
\newblock \bibinfo{title}{Variability in the location of high frequency
  oscillations during prolonged intracranial {EEG} recordings}.
\newblock \bibinfo{journal}{Nature Communications}
  \bibinfo{year}{2018};\bibinfo{volume}{9}(\bibinfo{number}{1}):\bibinfo{pages}{2155}.
\newblock \DOIprefix\doi{10.1038/s41467-018-04549-2}.
\bibitem[{Gliske et~al.(2016)Gliske, Irwin, Davis, Sahaya, Chestek and
  Stacey}]{gliske2016a}
\bibinfo{author}{Gliske\xfnm[ S.V.]}, \bibinfo{author}{Irwin\xfnm[ Z.T.]},
  \bibinfo{author}{Davis\xfnm[ K.A.]}, \bibinfo{author}{Sahaya\xfnm[ K.]},
  \bibinfo{author}{Chestek\xfnm[ C.]}, \bibinfo{author}{Stacey\xfnm[ W.C.]}.
\newblock \bibinfo{title}{Universal automated high frequency oscillation
  detector for real-time, long term {EEG}}.
\newblock \bibinfo{journal}{Clinical Neurophysiology}
  \bibinfo{year}{2016};\bibinfo{volume}{127}(\bibinfo{number}{2}):\bibinfo{pages}{1057--1066}.
\newblock \DOIprefix\doi{10.1016/j.clinph.2015.07.016}.
\bibitem[{Gliske et~al.(2020)Gliske, Qin, Lau, Alvarado-Rojas, Salami, Zelmann
  and Stacey}]{gliske2020}
\bibinfo{author}{Gliske\xfnm[ S.V.]}, \bibinfo{author}{Qin\xfnm[ Z.A.]},
  \bibinfo{author}{Lau\xfnm[ K.]}, \bibinfo{author}{Alvarado-Rojas\xfnm[ C.]},
  \bibinfo{author}{Salami\xfnm[ P.]}, \bibinfo{author}{Zelmann\xfnm[ R.]},
  \bibinfo{author}{Stacey\xfnm[ W.C.]}.
\newblock \bibinfo{title}{Distinguishing false and true positive detections of
  high frequency oscillations}.
\newblock \bibinfo{journal}{Journal of Neural Engineering}
  \bibinfo{year}{2020};\DOIprefix\doi{10.1088/1741-2552/abb89b}.
\bibitem[{Hoecker and Kartvelishvili(1996)}]{hoecker1996}
\bibinfo{author}{Hoecker\xfnm[ A.]}, \bibinfo{author}{Kartvelishvili\xfnm[
  V.]}.
\newblock \bibinfo{title}{{SVD} approach to data unfolding}.
\newblock \bibinfo{journal}{Nuclear Instruments and Methods in Physics Research
  Section A: Accelerators, Spectrometers, Detectors and Associated Equipment}
  \bibinfo{year}{1996};\bibinfo{volume}{372}(\bibinfo{number}{3}):\bibinfo{pages}{469--481}.
\newblock \DOIprefix\doi{10.1016/0168-9002(95)01478-0}.
\bibitem[{Karoly et~al.(2018)Karoly, Goldenholz, Freestone, Moss, Grayden,
  Theodore and Cook}]{karoly2018}
\bibinfo{author}{Karoly\xfnm[ P.J.]}, \bibinfo{author}{Goldenholz\xfnm[ D.M.]},
  \bibinfo{author}{Freestone\xfnm[ D.R.]}, \bibinfo{author}{Moss\xfnm[ R.E.]},
  \bibinfo{author}{Grayden\xfnm[ D.B.]}, \bibinfo{author}{Theodore\xfnm[
  W.H.]}, \bibinfo{author}{Cook\xfnm[ M.J.]}.
\newblock \bibinfo{title}{Circadian and circaseptan rhythms in human epilepsy:
  a retrospective cohort study}.
\newblock \bibinfo{journal}{The Lancet Neurology}
  \bibinfo{year}{2018};\bibinfo{volume}{17}(\bibinfo{number}{11}):\bibinfo{pages}{977--985}.
\newblock \DOIprefix\doi{10.1016/S1474-4422(18)30274-6}.
\bibitem[{Klerman et~al.(2017)Klerman, Wang, Phillips and
  Bianchi}]{klerman2017}
\bibinfo{author}{Klerman\xfnm[ E.B.]}, \bibinfo{author}{Wang\xfnm[ W.]},
  \bibinfo{author}{Phillips\xfnm[ A.J.K.]}, \bibinfo{author}{Bianchi\xfnm[
  M.T.]}.
\newblock \bibinfo{title}{Statistics for sleep and biological rhythms research:
  Longitudinal analysis of biological rhythms data}.
\newblock \bibinfo{journal}{Journal of Biological Rhythms}
  \bibinfo{year}{2017};\bibinfo{volume}{32}(\bibinfo{number}{1}):\bibinfo{pages}{18--25}.
\newblock \DOIprefix\doi{10.1177/0748730416670051}.
\bibitem[{Polyanin and Manzhirov(2008)}]{polyanin2008}
\bibinfo{author}{Polyanin\xfnm[ A.D.]}, \bibinfo{author}{Manzhirov\xfnm[
  A.V.]}.
\newblock \bibinfo{title}{Handbook of integral equations}.
\newblock \bibinfo{publisher}{Chapman and Hall/CRC}, \bibinfo{year}{2008}.
\bibitem[{Worrell et~al.(2004)Worrell, Parish, Cranstoun, Jonas, Baltuch and
  Litt}]{worrell2004}
\bibinfo{author}{Worrell\xfnm[ G.A.]}, \bibinfo{author}{Parish\xfnm[ L.]},
  \bibinfo{author}{Cranstoun\xfnm[ S.D.]}, \bibinfo{author}{Jonas\xfnm[ R.]},
  \bibinfo{author}{Baltuch\xfnm[ G.]}, \bibinfo{author}{Litt\xfnm[ B.]}.
\newblock \bibinfo{title}{High-frequency oscillations and seizure generation in
  neocortical epilepsy}.
\newblock \bibinfo{journal}{Brain}
  \bibinfo{year}{2004};\bibinfo{volume}{127}(\bibinfo{number}{7}):\bibinfo{pages}{1496--1506}.
\newblock \DOIprefix\doi{10.1093/brain/awh149}.

\end{thebibliography}

\end{document}


\setcounter{section}{8}
\section{Supplement: Derivation of Confidence Intervals}

\subsection{Background Information}

We note the variance on direct Monte Carlo integration (Eq. \ref{eq:Ihat}) is 
\begin{equation}
  \sigma^2_{\mathcal I} = \frac{1}{n-1} \sum_{k=1}^n \left[ g\left({\bm x}^{(k)}\right) - \hat{\mathcal I}\, \right]^2
\end{equation}
\citep{cappe2005} in the case of one basis function.  When using multiple basis functions, we have
\begin{eqnarray}
  \hat{\mathcal I}_i    &=& \frac{1}{n} \sum_{k=1}^n g_i\left({\bm x}^{(k)}\right),\\
  \left(C_{\hat{\bm{\mathcal{I}}}}\right)_{i,j} &=&
     \frac{1}{n-1} \sum_{k=1}^n
     \left[ g_i\left({\bm x}^{(k)}\right) - \hat{\mathcal I}_i \right]
     \left[ g_j\left({\bm x}^{(k)}\right) - \hat{\mathcal I}_j \right].
\end{eqnarray}

\subsection{The Unfolding Procedure}

\subsubsection{General Basis}

The covariance on $\beta$, Eq. \ref{eq:beta_estimate} is 
\begin{equation}
  \left(C_{\bm \beta}\right)_{i,j} = \frac{1}{n-1} \sum_{k=1}^n \left[f_i\left(\xM^{(k)}\right) - \beta_i\right]\left[f_j\left(\xM^{(k)}\right) - \beta_j\right].
\end{equation}
The covariance of $S$, Eq. \ref{eq:S_estimate}, is the 4-form
\begin{multline}
  \left(C_S\right)_{i,j;i^\prime,j^\prime} = \frac{1}{m-1} \\
     \times \sum_{k=1}^m
     \left[ V f_i\left(\xM^{\prime (k)}\right) f_j\left(\xT^{\prime (k)}\right) - S_{i,j}\right]
     \left[ V f_{i^\prime}\left(\xM^{\prime (k)}\right) f_{j^\prime}\left(\xT^{\prime (k)}\right) - S_{i^\prime,j^\prime}\right].
\end{multline}

Using first-order propagation of covariance, we can estimate the covariance on $\bm \alpha$, Eq. \ref{eq:alpha_formal}.  Formally, it is
\begin{equation}
  \label{eq:Calpha1}
    C_{\bm \alpha} = J_{\bm \beta} C_{\bm \beta} J_{\bm \beta}^T + J_S C_S J_S^T,
\end{equation}
with Jacobians defined as 
\begin{eqnarray}
  \left(J_{\bm \beta}\right)_{i,j} &=& \frac{\partial \alpha_i}{\partial \beta_j},\\
  \left(J_S\right)_{i,j,k} &=& \frac{\partial \alpha_i}{\partial S_{j,k}}.
\end{eqnarray}
The first Jacobian $J_{\bm \beta}$ is simply
\begin{equation}
  \label{eq:Jbeta}
  J_{\bm \beta} = S^{-1},
\end{equation}
based on Eq. \ref{eq:alpha_formal}. The second Jacobian $J_S$ can be obtained by implicit differentiation of Eq. \ref{eq:linear_eq}, which we write out in component form as
\begin{equation}
  \label{eq:component}
  \beta_k = \sum_{\ell=1}^N S_{k,\ell} \alpha_\ell.
\end{equation}
We then apply the operator $\frac{\partial}{\partial S_{i,j}}$ to both sides of Eq. \ref{eq:component}.
Noting the left side $\left(\frac{\partial \beta_k}{\partial S_{i,j}}\right)$ is zero, we have
\begin{eqnarray}
  0 &=& \sum_{\ell=1}^N \frac{\partial}{\partial S_{j,k}} \left( S_{k\ell}\alpha_\ell \right),\\
  &=& \sum_{\ell=1}^N \left( \frac{\partial S_{k,\ell}}{\partial S_{j,k}}\alpha_\ell + S_{k,\ell}\frac{\partial \alpha_\ell}{\partial S_{j,k}} \right),\\
  &=& \delta_{i,k} \alpha_j + \sum_{\ell=1}^N S_{k,\ell} \frac{\partial \alpha_\ell}{\partial S_{i,j}}.
\end{eqnarray}
Assuming that $S$ is invertible, yields
\begin{eqnarray}
  \left(J_{\bm S}\right)_{i,j,k}
     &=& \frac{\partial \alpha_i}{\partial S_{j,k}},\\
  \label{eq:JS}
     &=& -\left(S^{-1}\right)_{i,j}\alpha_k.
\end{eqnarray}
Lastly, for convenience we define matrix $C^\prime_{S}$ as
\begin{equation}
  \label{eq:CSprime}
  \left(C^\prime_S\right)_{j,j^\prime} = \sum_{k=1}^N \sum_{k^\prime=1}^N \alpha_k \left(C_S\right)_{j,k;j^\prime,k^\prime} \alpha_{k^\prime}.
\end{equation}
Substituting Equations \ref{eq:Jbeta}, \ref{eq:JS} and
\ref{eq:CSprime} into Eq. \ref{eq:Calpha1} yields the final equation
for the covariance on $\bm \alpha$,
\begin{eqnarray}
  C_{\bm \alpha} &=& S^{-1} C_{\bm \beta} S^{-T} + S^{-1} C_S^\prime S^{-T},\\
  \label{eq:Calpha2}
            &=& S^{-1}\left(C_{\bm \beta}+C^\prime_S\right)S^{-T}.
\end{eqnarray}

\subsubsection{Fourier Basis}

Recall $\bm \alpha^\prime$ is the normalized version of $\bm\alpha$, Eq. \ref{eq:alphaprime}.
The Jacobian of this transformation is
\begin{eqnarray}
  \label{eq:Jalpha}
  \left(J_{\bm \alpha}\right)_{i,j} = \frac{\partial \alpha^\prime_i}{\partial \alpha_j}
  &=& \left\{
  \begin{array}{ll}
    {-\alpha_i}/{2\alpha_0^2} & j=0,\\
    {1}/{2\alpha_0}      & i=j,\\
    0 & i \ne j.
  \end{array}
  \right.
\end{eqnarray}
The final covariance on $\bm \alpha^\prime$ is then
\begin{equation}
  \label{eq:Calphaprime}
  C_{\bm \alpha^\prime} = J_{\bm \alpha}\, C_{\bm \alpha}\, J_{\bm \alpha}^T,
\end{equation}
which can be directly computed from Equations \ref{eq:Calpha2} and \ref{eq:Jalpha}.

Lastly, we can transform the circular moments $\bm\alpha^\prime$ into magnitudes and phases, Eq. \ref{eq:mag}.
For notational convenience, we combine the magnitudes and phases into one vector $\bm\gamma$, with
\begin{equation}
  \gamma_k = \left\{\begin{array}{ll}
  |c_k|,                &1\le k \le N^\prime,\\
  \varphi_{k-N^\prime}, &N^\prime+1 \le k \le 2N^\prime.
  \end{array}\right.
\end{equation}
We again use first-order propagation of covariance, resulting in the covariance matrix for $\bm\gamma$ being
\begin{equation}
  \label{eq:Cgamma}
  C_{\bm\gamma} = J_{\bm\alpha^\prime}C_{\bm\alpha^\prime}J_{\bm\alpha^\prime}^T.
\end{equation}
The Jacobian matrix is
\begin{eqnarray}
  \left(J_{\bm\alpha^\prime}\right)_{i,j} &=& \frac{\partial \gamma_i}{\partial \alpha^\prime_j }\\
  &=& \left\{\begin{array}{ll}
  \frac{\alpha^\prime_i}{|c_i|}   \left( \delta_{i,j} + \delta_{i+N^\prime,j}\right) &  1 \le i \le N^\prime \\
  \frac{\alpha^\prime_j}{|c_i|^2} \delta_{i,j+N^\prime}                              &  N^\prime+1 \le i \le N \& 1 \le j \le N^\prime, \\ 
  \frac{\alpha^\prime_j}{|c_i|^2} \delta_{i,j+N^\prime}                              &  N^\prime+1 \le i \le N \& N^\prime+1 \le j \le N, \\ 
  \end{array}\right.
\end{eqnarray}